\documentclass[aps,prb,floatfix,twocolumn,showpacs]{revtex4-1}

\usepackage{graphicx}
\usepackage{color}
\usepackage{amsthm}
\usepackage{stmaryrd}
\usepackage{hyperref}

\begin{document}
\draft
\title{Spin Hall effect in graphene due to random Rashba field}
\author{A. Dyrda\l$^{1}$, J.~Barna\'s$^{1,2}$}

\address{$^1$Faculty of Physics, Adam Mickiewicz University,
ul. Umultowska 85, 61-614 Pozna\'n \\
$^2$  Institute of Molecular Physics, Polish Academy of Sciences,
ul. M. Smoluchowskiego 17, 60-179 Pozna\'n, Poland}
\date{\today }

\begin{abstract}
Spin Hall effect due to random Rashba spin-orbit coupling in the
two-dimensional honeycomb lattice of carbon atoms (graphen) is considered
theoretically. Using the Green function method and diagrammatic
technique we show that fluctuations of the Rashba
interaction around zero average value give rise to nonzero spin
Hall conductivity. Generally, the conductivity is not universal, but depends on the ratio of the
total momentum and spin-flip relaxation rates.
\end{abstract}

\maketitle

%%%%%%%%%%%%%%%%%%%%%%%%%%%%%%%%%%%%%%%%%%%%%%%%%%%%%%%%%%%%%%%%%%%%%%%%%%%%%%%%%%%%%%
%\section{Introduction} %%%%%%%%%%%%%%%%%%%%%%%%%%%%%%%%%%%%%%%%%%%%%%%%%%%%%%%%%%%%%%%
%%%%%%%%%%%%%%%%%%%%%%%%%%%%%%%%%%%%%%%%%%%%%%%%%%%%%%%%%%%%%%%%%%%%%%%%%%%%%%%%%%%%%%

{\it Introduction:} Interest in the phenomena due to spin-orbit coupling has revived
in recent years owing to the possibility of pure electrical
control of the spin degree of freedom. One of such phenomena is
the spin Hall effect (SHE), predicted by Dyakonov and Perel in
1971~\cite{dyakonov,dyakonovlett,hirsch}. The effect consists in a
spin current (or spin accumulation) flowing perpendicularly to an
external electric field in systems with strong
spin-orbit interaction. Of particular interest is the SHE in
nonmagnetic systems, where the induced spin current is
not accompanied by a charge current. Physical
mechanisms that lead to SHE are either of intrinsic or extrinsic
origin (for review see refs
\onlinecite{schlieman,engelrashba,dyakonov2008,vignale2010}). The
intrinsic SHE is associated with band structure modified
by an intrinsic spin-orbit coupling of the material under
consideration, and is a consequence of an unusual trajectory of
electrons in the momentum space. Extrinsic contribution to the SHE,
in turn, is a consequence of spin-orbit
interaction due to impurities and other structural defects
that lead to spin-dependent scattering  -- skew scattering and
side jump.

In a general case, spin-orbit coupling in a system contains
regular (spatially uniform or periodic) and random
components\cite{glazov2010}. Local imperfections of the system, like
random distribution of donors or impurities, may lead to local
enhancement/suppression  of the regular spin-orbit coupling and as a consequence charge
carriers in the system propagate  in an effective fluctuating
spin-orbit field. Such fluctuations of spin-orbit field strongly
modify spin relaxation and significantly affect spin
transport. Dugaev et al~\cite{dugaev2010} reported recently that
the random Rashba field  in two-dimensional electron gas generates nonzero SHE, even in the presence of potential
scattering by impurities. This is distinct from the well known
result for a two-dimensional electron gas with constant Rashba
coupling, where potential scattering by impurities
cancels out the universal intrinsic spin Hall
conductivity\cite{sinova,inoue}.

In this paper we consider SHE created by fluctuating Rashba
spin-orbit coupling in graphene --  a two-dimensional honeycomb
lattice of carbon atoms~\cite{Geim2007,katsnelson}. The
corresponding  intrinsic spin-orbit coupling in graphene is very
weak~\cite{gmitra2009}, so the quantized value of spin Hall
conductivity~\cite{kane}, that appears inside the spin-orbit energy gap, is
extremely difficult to be verified experimentally. In turn, the
Rashba type spin-orbit interaction in graphene appears generally
due to a substrate and can be relatively strong. This interaction includes usually a regular component, but a random
term may also appear due to ripples of the  graphene plane, disorder,
electron-phonon coupling in the substrate, presence of adatoms,
etc~\cite{huertas2007,ertler,zhang2012,Jiang2012}. The influence of
fluctuating Rashba field on spin relaxation in graphene has been
analyzed for two different models of the Rashba field fluctuations~\cite{zhang2012,dugaev2011}. Here we show that the spin
Hall conductivity associated with such fluctuations
is not universal and depends on the ratio of total momentum and spin relaxation times. We
consider two different types of the correlation function for the
spin-orbit coupling strength.
Below,  we briefly
describe the model under consideration and the theoretical method
used to calculate spin Hall conductivity. Then, we calculate
the Feynman diagrams representing various contributions to the
conductivity and also evaluate the spin dephasing time.

{\it Model and method:} Since the intrinsic spin-orbit interaction in a free standing
graphene is very small, it will be neglected in the following.
Accordingly, the low-energy electronic states around the Dirac
point $K$ are described by the
Hamiltonian~\cite{kane} $H^{K}_{0}$, which in the sublattice space
takes the matrix form
\begin{equation}
H^{K}_{0} = \left(
      \begin{array}{cc}
        0 & v (k_{x} - i k_{y}) \\
        v (k_{x} + i k_{y}) & 0 \\
      \end{array}
    \right)\, ,
\end{equation}
where $k_{x(y)}$ are the wavevector components, and $v=\hbar v_F$ with $v_F$ being the electron Fermi velocity.

As already mentioned above, an important spin-orbit interaction in graphene is
the Rashba coupling due to a substrate -- though
a related interaction may also appear in a free standing graphene with ripples. In the following we consider the case when the Rashba coupling
vanishes on average, but the corresponding coupling parameter
fluctuates in space around zero. These fluctuations of the Rashba field will
be treated as a perturbation, so the full Hamiltonian for the
point $K$ can be written as $H^{K} = H^{K}_{0} + V^{K}_{\mathbf{k,
k'}}$, where the perturbation $V^{K}_{\mathbf{k, k'}}$ takes the
form
\begin{equation}
V^{K}_{\mathbf{k, k'}}= \left(
                \begin{array}{cc}
                  0 & \lambda_{\mathbf{k}\,\mathbf{k}'} (\sigma_{y} + i \sigma_{x}) \\
                  \lambda_{\mathbf{k}\,\mathbf{k}'} (\sigma_{y} - i \sigma_{x})  & 0 \\
                \end{array}
              \right)\;,
\end{equation}
with  $\lambda_{\mathbf{k}\,\mathbf{k}'}$ describing the
corresponding Rashba coupling term, and $\sigma_{\alpha}$ ($\alpha
=x,y,z$) being the Pauli matrices in spin space. According to our
assumptions, $\langle \lambda_{\mathbf{q}}
\rangle = 0$ and $\langle \lambda_{\mathbf{q}}^{2} \rangle =
C_{\mathbf{q}} \neq 0$ for $\mathbf{q} = \mathbf{k} -
\mathbf{k'}$.

Following the Kubo-Streda formula~\cite{streda,yang}, we write the spin Hall
conductivity in the form
\begin{equation}
\label{p10} \sigma^{s_{z}}_{xy} = \sigma^{s_{z}\, I}_{xy} +
\sigma^{s_{z}\, II}_{xy}.
\end{equation}
The term $\sigma^{s_{z}\, II}_{xy}$ is the so-called topological
contribution to the spin Hall conductivity, which comes from states below the
Fermi level. This contribution, however, is equal to zero for the
model under consideration ($\sigma^{s_{z}\, II}_{xy} \neq 0$ only
inside the energy gap induced by intrinsic spin-orbit interaction,
which has been neglected here). In turn, the term $\sigma^{s_{z}\,
I}_{xy}$ is determined by states at the Fermi level and is given
by the formula
\begin{equation}
\label{p12} \sigma^{s_{z}\, I}_{xy} = \frac{e \hbar}{2 \pi} \int
\frac{d^{2} \mathbf{k}}{(2 \pi)^{2}} \mathrm{Tr} \{
\hat{j}^{s_{z}}_{x} G^{R}(\varepsilon_{F}) \hat{v}_{y}
G^{A}(\varepsilon_{F})\},
\end{equation}
where $G^{R(A)}_{\mathbf{k}}(\varepsilon_{F})$ is the retarded
(advanced) Green function at the Fermi level $\varepsilon_{F}$;
$\hat{j}^{s_{z}}_{x}=(1/2) \left[ v_{x}, s_{z}\right]_{+}$ is the
spin current density operator, and $v_{i} = (1/\hbar )( \partial H
/\partial k_{i})$ is the velocity operator. In the model under
consideration, $\hat{j}^{s_{z}}_{x}$ and $v_{y}$ take the
following explicit forms:
\begin{equation}
j^{s_{z}}_{x} = \frac{v}{2} \left(
                              \begin{array}{cc}
                                0 & \sigma_{z} \\
                                \sigma_{z} & 0 \\
                              \end{array}
                            \right)
\end{equation}
and
\begin{eqnarray}
v_{y} = i \frac{v}{\hbar} \left(
                          \begin{array}{cc}
                            0 & - \sigma_{0} \\
                            \sigma_{0} & 0 \\
                          \end{array}
                        \right),
\end{eqnarray}
where $\sigma_0$ is a unit matrix in the spin space.

{\it Spin Hall conductivity and spin relaxation:} Perturbation expansion for the Green functions in Eq.(4) leads to
a series of Feynman diagrams contributing to the spin Hall
conductivity (see Fig.1). Thus, one can write
\begin{eqnarray}
\sigma^{s_{z}}_{xy} = \frac{e \hbar}{2 \pi}
{\mathrm{Tr}}\sum_{\mathbf{k\,k'}}\sum_{n=1}^4 D_{n} \equiv
\sigma^{s_{z} (1)}_{xy}+ \sigma^{s_{z}(2)}_{xy} +
\sigma^{s_{z}(3+4)}_{xy}. \hspace{0.4cm}
\end{eqnarray}

The term $\sigma^{s_{z}(1)}_{xy}$ represents a contribution from
the bare bubble diagram and is generally a part of intrinsic spin
Hall conductivity. Since the nonperturbative part of the
Hamiltonian does not contain any term associated with spin-orbit
interaction, this term vanishes, $\sigma^{s_{z}(1)}_{xy} = 0$. The
second term in Eq.(7) can be written as
\begin{eqnarray}
\sigma^{s_{z}(2)}_{xy} =\frac{e \hbar}{2 \pi}\int \frac{d^{2}
\mathbf{k}}{(2 \pi)^{2}} \frac{d^{2} \mathbf{k'}}{(2
\pi)^{2}}\hspace{2.5cm}\nonumber\\ \times {\mathrm{Tr}}\left[
j^{s_{z}}_{x}G^{R}_{\mathbf{k}}V_{\mathbf{k
k'}}G^{R}_{\mathbf{k'}}v_{y}G^{A}_{\mathbf{k'}}V_{\mathbf{k'k}}G^{A}_{\mathbf{k}}\right].
\end{eqnarray}
In turn, the last term of Eq.(7), corresponding to the bottom
diagrams in Fig.1, takes the form
\begin{eqnarray}
\sigma^{s_{z}(3+4)}_{xy} =\frac{e \hbar}{2 \pi}\int \frac{d^{2} \mathbf{k}}{(2 \pi)^{2}} \frac{d^{2} \mathbf{k'}}{(2 \pi)^{2}}\hspace{2.5cm}\nonumber\\ \times {\mathrm{Tr}}\left[ j^{s_{z}}_{x}G^{R}_{\mathbf{k}}v_{y}G^{A}_{\mathbf{k}}V_{\mathbf{k k'}}G^{A}_{\mathbf{k'}}V_{\mathbf{k'k}}G^{A}_{\mathbf{k}}\right.\hspace{0.5cm}\nonumber\\
\left.\hspace{1.7cm}+ j^{s_{z}}_{x}G^{R}_{\mathbf{k}}V_{\mathbf{k k'}}G^{R}_{\mathbf{k'}}V_{\mathbf{k' k}}G^{R}_{\mathbf{k}}v_{y}G^{A}_{\mathbf{k}}\right].
\end{eqnarray}

\begin{figure}[t]
\centering \includegraphics[width=245pt]{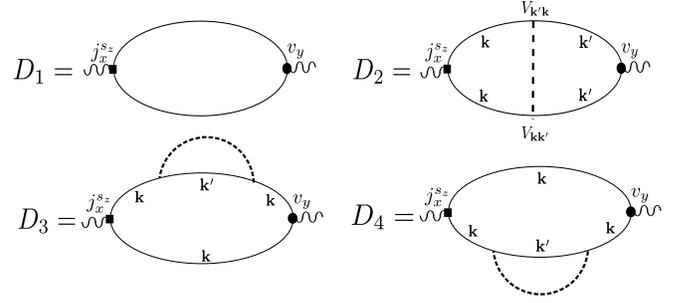}
  \caption{Feynman diagrams corresponding to various contributions to
  the spin Hall conductivity in the dc limit.}
\end{figure}

The Green functions $G^{R(A)}_{\mathbf{k}}$  have the form,
$G^{R(A)}_{\mathbf{k}} = \left[ (\varepsilon_F \pm i \Gamma) -
H^{K}_{0}\right]^{-1}$, where $\Gamma = \hbar/2\tau$. We assume
that $\tau$ is the total relaxation time, which  includes
scattering from impurities, $\tau_{i}$, and scattering from random
Rashba field, $\tau_{s}$. Thus one can write $1/\tau = 1/\tau_{i}
+ 1/\tau_{s}$. Note that for small impurity density the scattering
from random Rashba field is the dominant mechanism of relaxation
($\tau \rightarrow \tau_{s}$). To obtain $\tau_{s}$ we
calculate the self energy,
\begin{equation}
\Sigma^{R} = \int \frac{d^{2} \mathbf{k'}}{(2 \pi)^{2}}
V_{\mathbf{kk'}}G^{0R}_{\mathbf{k'}}V_{\mathbf{k' k}}.
\end{equation}
Taking into account the explicit form of the Green function
$G^{0R}_{\mathbf{k'}}$ corresponding to the Hamiltonian (1), one
obtains
\begin{equation}
\Sigma^{R} = 4 \varepsilon_F M \int \frac{d^{2} \mathbf{k'}}{(2
\pi)^{2}} \frac{\langle \lambda^{2}_{\mathbf{k k'}} \rangle
}{(\varepsilon_F - E_{1k'} + i0^{+})(\varepsilon_F - E_{2k'} +
i0^{+})},
\end{equation}
where $E_{1,2} = \pm vk \equiv \pm \varepsilon_{\bf k}$, and
\begin{equation}
M = \left(
                             \begin{array}{cc}
                               \frac{1}{2}(\sigma_{0} - \sigma_{z}) & 0 \\
                               0 & \frac{1}{2}(\sigma_{0} + \sigma_{z}) \\
                             \end{array}
                           \right).
\end{equation}
>From the above  equations follows that the self energy can be
written as
\begin{eqnarray}
\Sigma^{R} = - i \frac{M}{2 \pi}  \int dq\; q \int d\theta \;C_{q}\; \delta(\varepsilon_F - \varepsilon_{\mathbf{k - q}})\nonumber\\
= - i M \frac{k_{F}}{\pi v} \int^{2 k_{F}}_0 dq \frac{2 C_{q}}{\sqrt{4 k_F^{2} - q^{2}}}, \nonumber
\end{eqnarray}
or equivalently
\begin{equation}
\Sigma^{R} = - i \Gamma_{s} M,
\end{equation}
where $\Gamma_{s} = \hbar/2 \tau_{s}$. From Eq.(13) follows that
the explicit form of $\tau_{s}$ depends on the form of the
correlator $C_{q}$\cite{zhang2012,dugaev2011}.

Let us go now back to Eq.(8), which can be rewritten as
\begin{eqnarray}
\sigma^{s_{z}(2)}_{xy} =\frac{e}{2 \pi}\int \frac{d^{2}
\mathbf{k}}{(2 \pi)^{2}} \frac{d^{2} \mathbf{k'}}{(2
\pi)^{2}}\hspace{2.5cm}\nonumber\\\frac{16 |\lambda_{\mathbf{k
k'}}|^2 v^4 (k_{x} k^{\prime}_{x}-k_{y} k^{\prime}_{y})
\varepsilon_F  \Gamma
}{\left(\varepsilon^{2}_{\mathbf{k}}-(\varepsilon_F -i \Gamma
)^2\right) \left(\varepsilon^{2}_{\mathbf{k'}}-(\varepsilon_F -i
\Gamma )^2\right)}\nonumber\\ \times
\frac{1}{\left(\varepsilon^{2}_{\mathbf{k}}-(\varepsilon_F +i
\Gamma )^2\right)
\left(\varepsilon^{2}_{\mathbf{k'}}-(\varepsilon_F +i \Gamma
)^2\right)}.
\end{eqnarray}
Evaluation of this formula gives $\sigma^{s_{z}(2)}_{xy} =
0$. Thus, the spin Hall conductivity is given by the
bottom diagrams in Fig.1, $\sigma^{s_{z}}_{xy} = \sigma^{s_{z}(3 + 4)}_{xy}$, and for weak disorder one finds 
\begin{eqnarray}
\sigma^{s_{z}}_{xy}
= \frac{e}{4 \pi^{2}} \frac{v^{2} \varepsilon_F}{\Gamma} \int dq
\; q \int dk \, k \nonumber\\ \times \int d\theta\, C_{q}\,
\delta(\varepsilon^{2}_{\mathbf{k}} - \varepsilon_F^{2})\,
\delta(\varepsilon^{2}_{\mathbf{k - q}} - \varepsilon_F^{2}).
\end{eqnarray}
Finally the  spin Hall conductivity can be written in the following simple
form:
\begin{equation}
\sigma^{s_{z}}_{xy} = \frac{e}{4 \pi^{2}}
\frac{\varepsilon_F}{\Gamma} \int^{2 k_{F}}_{0} dq
\frac{C_{q}}{v^{2} \sqrt{4 k^{2}_{F} - q^{2}}} .
\end{equation}

Taking into account the formula for spin relaxation time, and including contributions from both K-points, we may write
the formula for spin Hall conductivity in the
form,
\begin{equation}
\sigma^{s_{z}}_{xy} = \frac{e}{4 \pi} \frac{\tau}{\tau_{s}}.
\end{equation}
This formula clearly shows that SHE is generally not universal. Potential scattering
by impurities  reduces the spin Hall conductivity,
but does not suppress it to zero.

The spin Hall conductivity depends on the form of the correlator
$C_{q}$. Dugaev et al~\cite{dugaev2011} assumed the correlator in
the form
\begin{equation}
C^{(1)}_{q} = 2 \pi \langle \lambda^{2} \rangle R^{2} e^{-qR}.
\end{equation}
Such fluctuations may originate from the ripples or some
impurities that exist at the surface of graphene. Using this
correlator one obtains (for a single $K$ point)
\begin{equation}
\sigma^{s_{z}}_{xy} = \frac{e}{4 \Gamma v} \langle \lambda^{2}
\rangle k_{F} R^{2} \left( I_{0}(2k_{F}R) - L_{0}(2k_{F}R)\right),
\end{equation}
where $I_{0}$ and $L_{0}$ are the modified Bessel and Struve
functions of zeroth order, respectively. From this relation one
finds asimple formula in some limiting cases,
\begin{equation}
\sigma^{s_{z}}_{xy} = \frac{e}{4 \pi} \frac{\langle \lambda^{2} \rangle}{v \Gamma} R \left\{
                                                                                       \begin{array}{cc}
                                                                                        \pi k_{F} R &; k_{F}R \ll 1 \\
                                                                                        1 &;k_{F}R \gg 1  \\
                                                                                       \end{array}
                                                                                     \right.\,.
\end{equation}
\begin{figure}[t]
\centering \includegraphics[width=230pt]{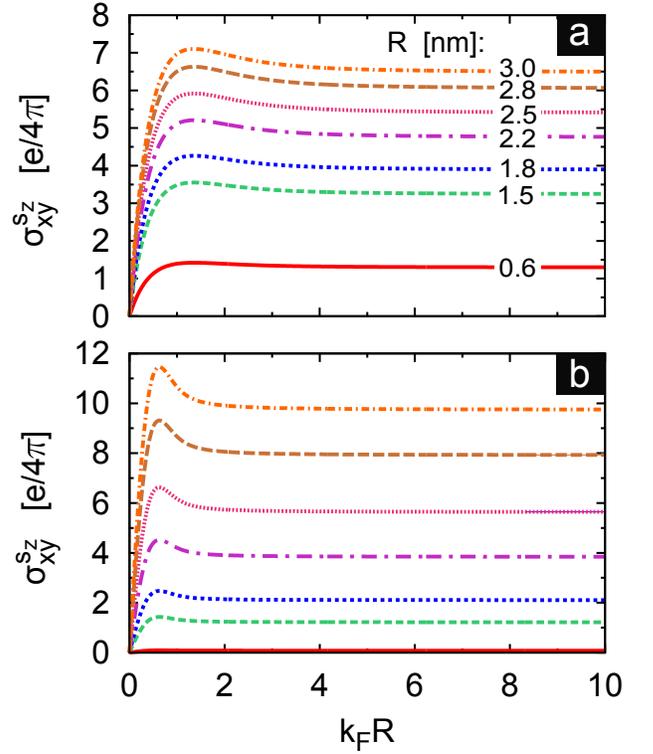}
 \caption{
Spin Hall conductivity as a function of $k_{F}R$ for constant (as
indicated) values of $R$. The parts a) and b) correspond to the
correlation function defined by Eq.(17) and Eq.(20), respectively.
Contributions from both $K$ points are included. The parameters used in the calculations are as follows:
$\langle \lambda^{2} \rangle =  25 \times 10^{-9}$  eV$^2$,
$v = 3.516 \times 10^{-10}$  eVm,
$n = 3\times10^{16}$ m$^{-2}$, and
$\Gamma = 6.58 \times 10^{-8}$ eV.
 }
\end{figure}

\begin{figure}[t]
\centering \includegraphics[width=230pt]{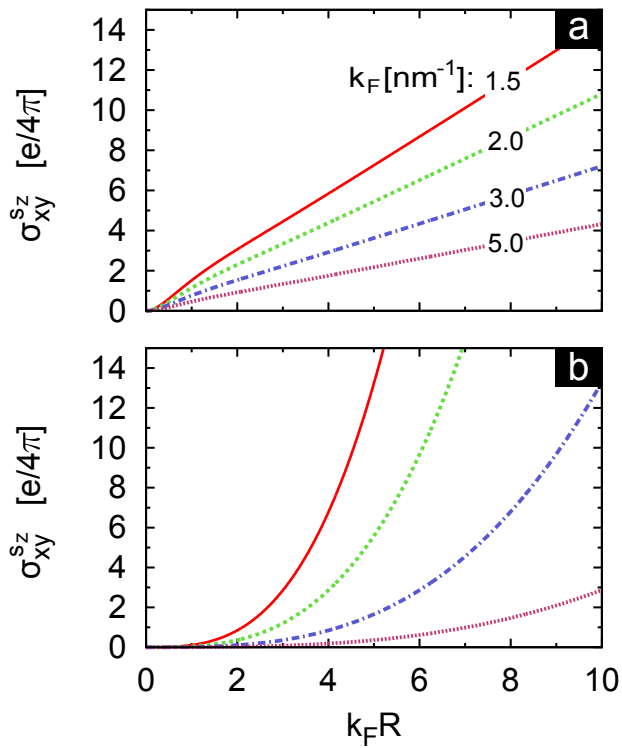}
  \caption{
Spin Hall conductivity as a function of $k_{F} R$ for constant (as
indicated) values of $k_F$. The parts a) and b) correspond to the
correlation function defined by Eq.(17) and Eq.(20), respectively.
Contributions from both $K$ points are included. The other
parameters are as in Fig.2.}
\end{figure}

In turn, Zhang and Wu~\cite{zhang2012} have proposed recently
another form of the correlator that may describe fluctuations of the
Rashba field due to the presence of adatoms distributed on top of
the graphene  surface and between the  graphene layer  and
substrate,
\begin{equation}
C^{(2)}_{q} = 4 \pi^{2} n \langle \lambda^{2} \rangle R^{4} e^{-q^{2}R^{2}}.
\end{equation}
where $n$ is the impurity density. The corresponding spin Hall conductivity (for single $K$ point)
takes the form
\begin{equation}
\sigma^{s_{z}}_{xy} = \frac{e n}{2 v \Gamma} \langle
\lambda^{2} \rangle k_{F} R^{4} \pi I_{0}(2k^{2}_{F}R^{2})
e^{-2k^{2}_{F}R^{2}},
\end{equation}
and in the limiting cases
\begin{equation}
\sigma^{s_{z}}_{xy} = \frac{e n}{2 v \Gamma} \langle \lambda^{2}
\rangle R^{3} \left\{
                                          \begin{array}{cc}
                                       \pi k_{F} R &; k_{F}R \ll 1 \\
                                         \frac{\sqrt{\pi}}{2} &;k_{F}R \gg 1  \\
                                        \end{array}
                                        \right.\,.
\end{equation}

{\it Numerical results:} Numerical results on the spin Hall conductivity in a general situation (arbitrary  $k_FR$) are presented in
Fig.2 and Fig.3. The contributions from both $K$ and $K^\prime$ points are
included there. Figure 2 shows the spin Hall conductivity as a
function of $k_FR$ for fixed values of the correlation length $R$,
as indicated. Parts (a) and (b) correspond to the two considered
forms of the correlator  $C_q$. In turn, Fig. 3 shows the spin
Hall conductivity as a function of $k_FR$ in the case when $k_F$
is fixed as indicated. As before, the upper and lower parts correspond
to the two forms of the correlator $C_q$.

From these two figures clearly follows that the spin Hall conductivity vanishes when
either the Fermi wavevector $k_F$ or the correlation length $R$
tend to zero. To understand this behavior one should bear in mind
that $k_F=0$ corresponds to the Fermi level at the Dirac points,
where the density of states vanishes. In turn, the limit of $R=0$
corresponds to the electron Fermi wavelength much longer than the
scale length of the disorder. These limiting values of the spin Hall copnductivity are in agreement
with the corresponding formulas (20) and (23). When $k_FR$ grows (keeping $R$ constant),
the conductivity initially grows with $k_FR$ and after reaching maximum it tends to the
limits for $k_FR \gg 1$ according to the formulas (20) and (23).
Note, these limits depend on $R$, as follows from (20) and (23).
In turn, when $k_F$ is fixed (see Fig.3), the spin Hall conductivity increases with
increasing  $k_FR$, but the rate of this increase depends on $k_F$.

To conclude, we have analyzed the spin Hall effect induced by fluctuating
Rashba spin-orbit interaction  in a monolayer graphene. The key
assumption was the absence of uniform Rashba term, i.e. the
assumption that the Rashba interaction vanishes on average. Two
different forms of the correlation function have been considered,
which may correspond to different physical situations. We note,
that such random Rashba interaction may appear as a result of
ripples in a free standing graphene or due to impurities on both
sides of the graphene plane.

It has been shown, that the spin Hall conductivity induced by the
fluctuating Rashba field is equal to $(e/4\pi )(\tau/\tau_s)$, independently on the form of the
correlator, and is generally not
universal. A universal value occurs only when both $\tau$ and
$\tau_s$ are equal (as in the absence of any other defects, except the fluctuating Rashba term).
It is thus evident that potential scattering  by defects suppresses the spin Hall effect, but this suppression is not complete, similarly as in the case of two-dimensional electron gas.

\subsection*{Acknowledgments}

This work has been supported in part by the European Union under
European Social Fund Operational Programme 'Human Capital' (POKL.04.01.01-00-133/09-00) and in part by National Science
Center (NCN, Poland) as a research project in years 2012-2014,
grant No. DEC-2011/03/N/ST3/02353. The authors also acknowledge
valuable  discussions with V.K. Dugaev and his  useful comments on the manuscript.

%%%%%%%%%%%%%%%%%%%%%%%%%%%%%%%%%%%%%%%%%%%%%%%%%%%%%%%%%%%%%%%%%%%%%%%%%%%%%%%%%%%%%%%%%%%%%%%%%%%%%%%%%

\end{document}